\documentclass[conference]{IEEEtran}
\usepackage{multirow}
\usepackage{booktabs}
\usepackage{amsfonts}
\usepackage{placeins}
\usepackage{array}
\usepackage{amsmath,cases,amssymb}
\usepackage{amsmath}
\usepackage{graphicx}
\usepackage{cite}
\usepackage{subfigure}
\usepackage{epsfig}
\usepackage{todonotes}
\usepackage{url}
\usepackage{algorithm}
\usepackage{algorithmic}
\usepackage{epstopdf}
\usepackage{color}
\usepackage{algorithm}
\DeclareGraphicsExtensions{.eps,.pdf}

\usepackage{scalerel}

\newcommand{\bseq}{\begin{subequations}}
\newcommand{\eseq}{\end{subequations}}
\newcommand{\baln}{\begin{align}}
\newcommand{\ealn}{\end{align}}
\newcommand{\balnd}{\begin{aligned}}
\newcommand{\ealnd}{\end{aligned}}
\newcommand{\beq}{\begin{equation}}
\newcommand{\eeq}{\end{equation}}
\newcommand{\beqn}{\begin{eqnarray}}
\newcommand{\eeqn}{\end{eqnarray}}
\newcommand{\beqno}{\begin{eqnarray*}}
\newcommand{\eeqno}{\end{eqnarray*}}
\newcommand{\bma}{\begin{displaymath}}
\newcommand{\ema}{\end{displaymath}}
\newcommand{\bnu}{\begin{enumerate}}
\newcommand{\enu}{\end{enumerate}}
\newcommand{\bce}{\begin{center}}
\newcommand{\ece}{\end{center}}
\newcommand{\btb}{\begin{tabular}}
\newcommand{\etb}{\end{tabular}}
\newcommand{\bIEEEeq}{\begin{IEEEeqnarray}}
\newcommand{\eIEEEeq}{\end{IEEEeqnarray}}

\newtheorem{theorem}{Theorem}

\newcommand{\st}{{\mathrm{s.t.}}}

\newcommand{\calK}{{\mathcal {K}}}
\newcommand{\calM}{{\mathcal {M}}}
\newcommand{\calN}{{\mathcal {N}}}

\newcommand{\calT}{{\mathcal {T}}}
\newcommand{\non}{\nonumber}
\newcommand{\subnum}{\IEEEyessubnumber}

\newcommand{\balpha}{\boldsymbol{\alpha}}

\newcommand{\bmu}{\boldsymbol{\mu}}

\newcommand{\BS}{\mathtt{BS}}
\newcommand{\LEO}{\mathtt{LEO}}
\newcommand{\UEk}{\mathtt{UE}_{k}}
\newcommand{\UE}{\mathtt{UE}}

\newcommand{\blambda}{\boldsymbol{\lambda}}

\newcommand{\bp}{\mathbf{p}}
\newcommand{\bP}{\mathbf{P}}

\newcommand{\gmnT}{g_{m,n}[t]}
\newcommand{\bW}{\mathbf{W}}

\newcommand{\Nsc}{N_{\mathtt{SC}}}
\newcommand{\Wsc}{W_{\mathtt{SC}}}
\newcommand{\calNsc}{\calN_{\mathtt{SC}}}

\allowdisplaybreaks

\title{Two-tier User Association and Resource Allocation Design for Integrated Satellite-Terrestrial Networks}

\vspace{-3mm}

\author{\IEEEauthorblockN{Hung Nguyen-Kha${}^{\dagger}$, Vu Nguyen Ha${}^{\dagger}$, Eva Lagunas${}^{\dagger}$, Symeon Chatzinotas${}^{\dagger}$, and Joel Grotz${}^{\ddagger}$}

\IEEEauthorblockA{\textit{${}^{\dagger}$Interdisciplinary Centre for Security, Reliability and Trust (SnT), University of Luxembourg, Luxembourg} \\
       \textit{${}^{\ddagger}$SES S.A., Luxembourg}}
       
       \vspace{-8mm}
       }

\begin{document}
\maketitle

\begin{abstract}
This paper presents a study of an integrated satellite-terrestrial network, where Low-Earth-Orbit (LEO) satellites are used to provide the backhaul link between base stations (BSs) and the core network. The mobility of LEO satellites raises the challenge of determining the optimal association between LEO satellites, BSs, and users (UEs). The goal is to satisfy the UE demand while ensuring load balance and optimizing the capacity of the serving link between the BS and the LEO satellite.
To tackle this complex optimization problem, which involves mixed-integer non-convex programming, we propose an iterative algorithm that leverages approximation and relaxation methods. The proposed solution aims to find the optimal two-tier satellite-BS-UE association, sub-channel assignment, power and bandwidth allocation in the shortest possible time, fulfilling the requirements of the integrated satellite-terrestrial network. 
\end{abstract}
%
\section{Introduction}


Low-Earth-Orbit (LEO) satellite constellations have been gaining attention in recent years as a promising solution for providing global connectivity as well as low-latency and high-capacity broadband services, compared to traditional geostationary satellites \cite{SatCom_survey_and_challenge, Servey_NGSO,VuHa_ICC23}. 
On the other hand, the satellite constellations can also help the current terrestrial networks deal with critical challenges on limited coverage of the rural, suburban, and city-edge areas \cite{ISTN_Toward_6G_App_challenge}. Even though there is a lot of interest and effort, it is still unclear if direct broadband communications between handheld devices and satellites can be achieved due to the terminal size and antenna limitations \cite{ISTN_Toward_6G_App_challenge}. Therefore, integrated satellite-terrestrial networks (ISTNs) are a potential solution to address the coverage issue in underserved areas. However, the design and optimization of ISTNs that utilize LEO satellites as a backhaul link poses significant challenges, particularly with regard to the association between LEO satellites, base stations (BSs), and users (UEs).

There has been a significant amount of research on the topic of satellite-terrestrial networks, including studies on the integration of LEO satellites, user association, and resource allocation \cite{Tedros_ICC23,twotier_UE_HAP_LEO_association, LEO_HO_Bipartite_GamePotential, LEO_5G_Offloading, LEO_IoT_MinTime, Hung_WSA23, UltraDenseLEO_Offload_TN}. 
In \cite{twotier_UE_HAP_LEO_association}, a matching algorithm was proposed to solve the association problem between UEs in a disaster area, high-altitude-platforms (HAPs), and the LEO satellite for computation offloading.
The author in \cite{LEO_HO_Bipartite_GamePotential} studied a handover solution between UEs and LEO satellites based on the bipartite graph and game potential.
In \cite{LEO_5G_Offloading}, the LEO-backhauled small cell is deployed to assist the traditional uplink TNs for data offloading, wherein the transmission selection of UEs and resource allocation problem was studied to maximize the sum-rate system.
In \cite{LEO_IoT_MinTime}, the authors considered a joint power allocation, sub-carrier assignment problem minimizing the completion time for IoT schemes, where IoT devices upload their data to an LEO satellite via center earth stations.
The authors in \cite{Hung_WSA23} regarded resource allocation and association between satellite UE/BS and LEO satellite design to minimize the transmit power. 
The authors in \cite{UltraDenseLEO_Offload_TN} studied a multi-objective optimization problem regarding user association, resource allocation, and service price setting for an ISTN where the TN can offload its users to the satellite network.
However, most of the existing literature has focused on the single-tier association between the satellite and BS or UE, and limited attention has been given to the two-tier association design \cite{LEO_HO_Bipartite_GamePotential, LEO_5G_Offloading, LEO_IoT_MinTime, Hung_WSA23, UltraDenseLEO_Offload_TN}. Moreover, jointly optimizing both resource allocation and user association in ISTNs with unbalanced backhaul capacity has not been fully addressed.

In this paper, we investigate the design of a two-tier user association and resource allocation for an integrated satellite-terrestrial network that utilizes LEO satellites as a backhaul link, wherein BSs decode-and-forward the received data from UEs to the LEO satellite. Our focus is on the optimization problem of satellite-BS-UE association and resource allocation under the constraints of load balance and UE demand. 
To begin with, we formulate an optimization that takes into account all these design aspects.
This problem considers the continuous variables corresponding to the bandwidth (BW) and power allocation and the binary variables related to the two-tier association mechanism, which classifies the problem as an NP-hard mixed integer non-linear programming (MINP).
The resulting problem is even more challenging due to the non-convex sum rate over two hops, users-base stations and base stations-satellites. 
To address this problem, we propose an iterative algorithm employing both compressed sensing and successive convex approximation methods. 
A greedy mechanism is also presented for comparison purposes. The numerical results provide valuable insights into the design and optimization of integrated satellite-terrestrial networks and can help to advance the development of next-generation communication systems.
\section{System Model and Problem Formulation}
Consider an integrated system consisting of $M$ LEO satellites, $N$ ground-based base stations (BSs), and $K$ uplink terrestrial users (UEs). The BSs provide radio access service to the UEs while the LEO satellite acts as the backhaul link between the BSs and the core network. For ease of reference, we define the sets of LEO satellites, BSs, and UEs as $\calM$, $\calN$, and $\calK$, respectively, and denote the $m$-th LEO satellite as $\LEO_{m}$, the $n$-th BS as $\BS_{n}$, and the $k$-th UE as $\UEk$. The system operates within a time window $T_W = N_T T_S$, where $N_T$ is the number of time-slots (TSs) and $T_S$ is the duration of each TS. The set of TSs is defined as $\calT_S$. The $K$ users require transmission of their respective data amounts $D_1,\dots,D_K$ bits, and the transmission model is outlined as follows.

\vspace{-2mm}

\subsection{Transmission from UE to BS}
\begin{figure}[!t]
	\centering
\includegraphics[width=50mm,height=25mm]{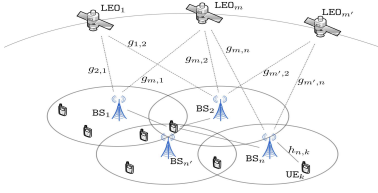}
	\caption{System model.}
	\label{fig:satbsue2tier}
 \vspace{-8mm}
\end{figure}
In the terrestrial network, UEs transmit their data to the BS, wherein each BS can serve multiple UEs while each UE can be served by at most one BS. For the UE located at the overlap area covered by more than one BS, it could be served by one of these BSs. 
In addition, the transmission bandwidth $W^\BS$ for each BS is divided into  $ \Nsc $ sub-channels (SCs) with the bandwidth of $\Wsc$, which are allocated to served UEs. Let $\calNsc \triangleq \{1,\dots,\Nsc\}$ be the set of SCs. To perform the BS-UE association and the SC allocation at TS $t$, a new variable $ \balpha [t] \triangleq [\alpha_{n,k,s}[t]]_{\forall (n,k,s) \in (\calN \times \calK \times \calNsc)} $ is introduced as $\alpha_{n,k,s}[t]=1$ if $\BS_{n}$ serves $\UEk$  over SC $s$ at TS $t$, and $\alpha_{n,k,s}[t]=0$, otherwise.
One assumes that each SC can be assigned to at most one served UE at every BS; furthermore, each UE can be assigned to at most $ \bar{S}$ SCs at each TS due to the limited processing ability at the UEs. These yield the following constraints,
\vspace{-1mm}
\beqn
(C1): \scaleobj{.8}{\sum_{\forall k \in \calK} }\alpha_{n,k,s}[t] \leq 1, \forall (n,s) \in (\calN \times \calNsc), \forall t \in \calT_S, \\
(C2): \scaleobj{.8}{\sum_{\forall s \in \calNsc} }\alpha_{n,k,s}[t] \leq \bar{S}, \forall (n,k) \in (\calN \times \calK), \forall t \in \calT_S.
\eeqn
Additionally, every UE is assigned to only one BS, which results in the following constraint,
\vspace{-1mm}
\bIEEEeq{ll}
(C3): \scaleobj{.8}{\sum_{\forall n }} \Big\Vert \scaleobj{.8}{\sum_{\forall s}} \alpha_{n,k,s}[t] \Big\Vert_0 \leq 1, \forall k, \forall t \in \calT_S. \eIEEEeq
Let $ p_{n,k,s}[t] $ be the transmitted power of $ \UEk $ over SC $s$ to $ \BS_n $ and $\bp[t] \triangleq [p_{n,k,s}[t]]_{\forall (n,k,s)}$. The received signal at $\BS_{n}$ over SC $s$ in TS $t$ can be expressed as
$y_{n,s}^{\BS,t} (\bp[t],\balpha[t]) \! = \!\! \scaleobj{.8}{\sum_{\forall k \in \calK}} \sqrt{\alpha_{n,k,s}[t] p_{n,k,s}[t] } \bar{h}_{n,k,s}[t] s_k \! + \! n_n,$
where $\bar{h}_{n,k,s}[t]$ is the channel coefficient between $\UEk$ and $\BS_{n}$, $s_k[t]$ with $\mathbb{E}\{|s_k[t]|^2 \}=1$ is the transmitted symbol of $\UEk$ in TS $t$, and $n_n \sim \mathcal{CN}(0,\sigma_n^2)$ is the additive Gaussian noise. 
Then, if $\UEk$ is served by $\BS_{n}$ and assigned SC $s$, the SINR corresponding to the data transmission of $\UEk$ over SC $s$ can be written as
\vspace{-1mm}
\begin{IEEEeqnarray}{lcl} 
\label{eq: SINR UEk at BSn SC s}
\mathtt{SINR}_{n,k,s}^t & := &	\gamma^{\UE,t}_{n,k,s}(\bp[t],\balpha[t]) \nonumber \\
& = & \scaleobj{1.2}{\frac{ \alpha_{n,k,s}[t] p_{n,k,s}[t] h_{n,k,s}[t] }{\sum_{\forall j \neq k}{h_{n,j,s}[t] \left(\sum_{\forall i} \alpha_{i,j,s}[t] p_{i,j,s}[t] \right)} + \sigma_n^2}}, \quad
\end{IEEEeqnarray}
where $ h_{n,k,s}[t] = |\bar{h}_{n,k,s}[t]|^2$ denotes the channel gain of link $ \UEk - \BS_{n} $ at SC $s$ in TS $ t $. Hence, the transmission data of $ \UEk $ over SC $s$ received at $\BS_{n}$ can be expressed as
\begin{IEEEeqnarray}{ll} \label{eq: rate UEk at BSn SC s}
	R_{n,k,s}^{\UE,t} (\bp[t],\balpha[t]) = T_S \Wsc \log_2(1+\gamma^{\UE,t}_{n,k,s}(\bp[t],\balpha[t])).
\end{IEEEeqnarray}
Then, the total achievable rate of $ \UEk $ in TS $ t $ is written as
\beq \label{eq: rate UEk}
R_k^{\UE,t} (\bp[t],\balpha[t]) = \scaleobj{.8}{ \sum_{\forall n \in \calN} \sum_{\forall s \in \calNsc} }  R_{n,k,s}^{\UE,t} (\bp[t],\balpha[t]).
\eeq
\vspace{-3mm}

\subsection{Transmission from BS to LEO}
\vspace{-1mm}
In this transmission stage, one assumes that the BSs upload all its received data to the core network through the LEO satellite in the different sub-channels, where the spectrum among LEO satellites are orthogonal. 
Regarding the association between BSs and LEOs, a new variable $ \bmu[t] \triangleq [\mu_{m,n}[t]]_{\forall (m,n)} $ presenting LEO-BS connection at TS $t$ is introduced as,
\vspace{-2mm}
\begin{align} \label{eq: BS-UE association}
	\mu_{m,n}[t]= 
	\begin{cases}
		1, &  \BS_{n} \text{ is served by } \LEO_{m} \text{ at TS } t, \\
		0 ,&  \text{otherwise}.
	\end{cases}
\end{align}
Note that each BS can be served by at most one LEO satellite at each TS, which is formed into the following constraint,
\vspace{-2mm}
\beq
(C4): \quad \scaleobj{.8}{\sum_{\forall m \in \calM} }\mu_{m,n}[t] \leq 1, \forall n \in \calN, 
\eeq
Denote $W_m^{\mathtt{LEO}}[t]$ as the maximum BW available at $\LEO_{m}$ in TS $t$ which can be utilized for the BS backhaul links. Let $W^\BS_{m,n} [t]$ be the BW of $ \LEO_{m} $ allocated to $\BS_{n}$ in TS $t$, we have
\beq
(C5): \scaleobj{.8}{\sum_{\forall n \in \calN} }\mu_{m,n}[t] W^\BS_{m,n} [t] \leq W_m^{\mathtt{LEO}}[t], \forall (m,t).
\eeq
Assuming that $\BS_{n}$ is served by $\LEO_{m}$, the transmission data of $ \BS_{n} $ at $ \LEO_{m} $ in TS $t$ is expressed as
\begin{IEEEeqnarray}{ll} \label{eq: rate BSn at LEOm}
	R_{m,n}^{\BS,t} (\bP[t],\bW^\BS[t] ,\bmu[t]) \nonumber \\ \hspace{0.5cm} =  \mu_{m,n}[t] T_S W^\BS_{m,n} [t] \log_2 \Big(1+ \gamma^{\BS,\mathtt{t}}_{m,n}(\bP[t],\bW^\BS [t])\Big), \hspace{0.5cm}
\end{IEEEeqnarray}
where $\gamma^{\BS,\mathtt{t}}_{m,n}(\bP[t],\bW^\BS [t])={\frac{ P_{n}[t] \gmnT}{W^\BS_n [t] \delta_m} }$ is the SNR of $\BS_n$ at $\LEO_m$ if they are associated, $\bP[t] \triangleq [P_n [t]]_{\forall n}$, $P_n [t]$ is the transmit power of $\BS_{n}$ and $ \gmnT $ is the channel gain between  $ \BS_{n} $ and $ \LEO_{m} $ in TS $t$, $\bW^\BS[t] \triangleq [W^\BS_{m,n} [t]]_{\forall (m,n)}$, and $\delta_m$ is the noise power at $\LEO_{m}$. Then, the transmission rate of $ \BS_{n} $ at TS $t$ is given as
\beq
\hspace{-2mm} R_n^{\BS,t} \! (\bP[t],\!\bW^\BS[t],\!\bmu[t]) = \!\!\! \scaleobj{.8}{ \sum_{\forall m \in \calM} } R_{m,n}^{\BS,t} (\bP[t],\! \bW^\BS[t],\bmu[t]).
\eeq
In order to successfully forward all the data from UEs associated to $\BS_{n}$, the following condition must be guaranteed,
\beq
\hspace{-4mm} (C6)\!\!: \!\! \scaleobj{.8}{\sum_{\forall (k,s)} } \!\! R_{n,k,s}^{\UE,t} \! (\bp[t],\! \balpha[t]) \!\! \leq \!\!
R_n^{\BS,t} \! (\bP[t], \!\! \bW^\BS[t],\! \bmu[t]), \! \forall (n,t). \!\!
\eeq

\subsection{Problem Formulation}
We aim to minimize the transmission time required for offloading all user demands to the core network in order to minimize the end-to-end latency. To do so, we first express the remaining data of $\UEk$ right after TS $t$ as
\beq \label{eq: data remain}
d_k[t] = \max\Big(0,D_k - \scaleobj{.8}{\sum_{u=1}^{t}} R_k^{\UE,u} (\bp[u],\balpha[u]) \Big).
\eeq
Then, the design objective is formulated as the TS-index minimization problem as follows.
\vspace{-2mm}
\begin{IEEEeqnarray}{cl}\label{eq: Min time}
	\hspace{-8mm} \min_{\bp,\bP,\bW^\BS,\balpha,\bmu,v}  \quad & v	\subnum	\label{eq: Min time a}	\\
	\hspace{-24mm} \st     & \hspace{-16mm} \text{constraints } (C1)-(C6), \nonumber \\
	& \hspace{-16mm} (C7): \scaleobj{.8}{\sum_{\forall n \in \calN} \sum_{\forall s \in \calNsc}} \alpha_{n,k,s}[t] p_{n,k,s}[t] \leq p_k^{\max}, \forall (k,t), \subnum \label{eq: Min time c}\\
	& \hspace{-16mm} (C8): P_n [t] \leq P_n^{\max}, \forall n \in \calN, \forall t \in \calT_S, \subnum \label{eq: Min time d}\\
	& \hspace{-16mm} (C9): d_k[v] = 0, \forall k \in \calK, \subnum \\
	& \hspace{-16mm} (C10): \alpha_{n,k,s}[t], \mu_{m,n}[t] \in \{0,1\}, \forall (m,n,k,s,t), \subnum  \label{eq: Min time e}
\end{IEEEeqnarray}
where $(C7)-(C8)$ stand for the limited power constraints.

\section{Proposed Solution}
It can be observed in \eqref{eq: Min time}, to satisfy the user demand in the shortest time, UEs having the large remaining data demand at each TS should be prior to being served. Therefore, we transform and derive problem \eqref{eq: Min time} to an equivalent optimization problem at TS $t$ \cite{Min_Time_To_Weight_Max_SR} as
\vspace{-1mm}
\bIEEEeq{cl}\label{eq: Max rate}
\max_{\bp,\bP,\bW^\BS,\balpha,\bmu}  & 	{\scaleobj{.8}{\sum_{\forall (k,n,s)}}} \omega_k[t] R_{n,k,s}^{\UE,t} (\bp[t],\balpha[t])	\nonumber \\
\st     & \text{constraints } (C1)-(C8),(C10),  
\eIEEEeq
where $\bp \triangleq \{\bp[t]\}_{\forall t},\bP \triangleq \{\bP[t]\}_{\forall t},\bW^\BS \triangleq \{\bW^\BS[t]\}_{\forall t},\balpha \triangleq \{\balpha[t]\}_{\forall t}$ and $\bmu \triangleq \{\bmu[t]\}_{\forall t}$; $\omega_k[t] = d_k[t]$ is the demanding weight of $\UEk$ at TS $t$. It can see that problem \eqref{eq: Max rate} is a mixed-integer non-convex programming, which is not trivial to be solved, owing to the coupling between binary and continuous variables and the non-convexity functions of transmission data.

\subsection{Compressed-Sensing Approach} 
If $ \UE_{k} $ is not served by $ \BS_{n} $ over SC $ s $, the corresponding transmit power over SC $s$ to $ \BS_{n} $ should be zero. Similarly, $W_m^{\mathtt{LEO}}[t] = 0$ implies that $\BS_{n}$ does not connect to $\LEO_{m}$. Accordingly, we have the following relationship constraints
\begin{IEEEeqnarray}{ll} \label{eq: norm0}
	\alpha_{n,k,s}[t] &= \big\Vert p_{n,k,s}[t] \big\Vert_0,  \quad \forall (n,k,s),\subnum \\
    \Big\Vert \scaleobj{.8}{\sum_{\forall s} }\alpha_{n,k,s}[t] \Big\Vert_0 &= \Big\Vert \scaleobj{.8}{\sum_{\forall s}} p_{n,k,s}[t] \Big\Vert_0, \quad \forall (n,k), \subnum \\
	\mu_{m,n}[t] &= \big\Vert W^\BS_{m,n}[t] \big\Vert_0,  \quad \forall (m,n). \subnum
\end{IEEEeqnarray}
Subsequently, the binary variables $\balpha$ and $\bmu$ in problem \eqref{eq: Max rate} can be performed by the continuous ones $\bp$ and $\bW$ thanks to \eqref{eq: norm0}, respectively. However, dealing with the sparsity issue of $\ell_0$ norm in \eqref{eq: norm0} is very challenging. To efficiently address this issue, we exploit the approximation method for re-weighted $\ell_1$ minimization as in \cite{compress_sensing,VuHa_TGCN20}. In particular, $\ell_0$-norm components in \eqref{eq: norm0} can be approximated at iteration $i$ as
\begin{IEEEeqnarray}{ll} \label{eq: norm1 reweight}
	\big\Vert p_{n,k,s}[t] \big\Vert_0 &= \zeta_{n,k,s}^{(i)}[t] p_{n,k,s}[t] ,  \quad \forall (n,k,s,t),\subnum \\
    \Big\Vert \scaleobj{.8}{\sum_{\forall s}} p_{n,k,s}[t] \Big\Vert_0 &= \xi_{n,k}^{(i)}[t] \scaleobj{.8}{\sum_{\forall s}} p_{n,k,s}[t], \quad \forall (n,k,t), \subnum \\
	\big\Vert W^\BS_{m,n}[t] \big\Vert_0 &= \chi_{m,n}^{(i)}[t] W^\BS_{m,n}[t],  \quad \forall (m,n,t), \subnum
\end{IEEEeqnarray}
where $\zeta_{n,k,s}^{(i)}[t], \xi_{n,k}^{(i)}[t]$ and $\chi_{m,n}^{(i)}[t]$ are the weights, which are updated as, $\zeta_{n,k,s}^{(i)}[t] =  {{1}/\scaleobj{.9}{(p_{n,k,s}^{(i-1)}[t] + \epsilon)}}$, $\xi_{n,k}^{(i)}[t] = {1}/\scaleobj{.9}{(\scaleobj{.8}{\sum_{\forall s}} p_{n,k,s}^{(i-1)}[t] + \epsilon)}$, and $\chi_{m,n}^{(i)}[t] = {1}/\scaleobj{.9}{(W^\BS_{m,n}[t]^{(i-1)} + \epsilon)}$
in which $\epsilon$ is a sufficiently small positive number.
Thanks to \eqref{eq: norm0} and \eqref{eq: norm1 reweight}, constraints $ (C1)$-$(C7) $ can be rewritten as
\begin{IEEEeqnarray}{ll}
\hspace{-5mm}	(\tilde{C}1):  \scaleobj{.8}{\sum_{\forall k \in \calK}} \zeta_{n,k,s}^{(i)}[t] p_{n,k,s}[t] \leq 1, \forall (n,s,t) , \subnum \\
\hspace{-5mm}	(\tilde{C}2):  \scaleobj{.8}{\sum_{\forall s \in \calNsc}} \zeta_{n,k,s}^{(i)}[t] p_{n,k,s}[t] \leq \bar{S}, \forall (n,k,t), \subnum \\
\hspace{-5mm}	(\tilde{C}3):  \scaleobj{.8}{\sum_{\forall n } }\xi_{n,k}^{(i)}[t] \sum_{\forall s} p_{n,k,s}[t] \leq 1, \forall (k,t), \subnum \\
\hspace{-5mm}	(\tilde{C}4):  \scaleobj{.8}{\sum_{\forall m \in \calM}} \chi_{m,n}^{(i)}[t] W^\BS_{m,n}[t] \leq 1, \forall n \in \calN, \subnum \\
\hspace{-5mm}	(\tilde{C}5):  \scaleobj{.8}{\sum_{\forall n \in \calN}}  W^\BS_{m,n} [t] \leq W_m^{\mathtt{LEO}}[t], \quad \forall (m,t) , \subnum \\
\hspace{-5mm}	(\tilde{C}6):  \scaleobj{.8}{\sum_{\forall (k,s)}} \!\! R_{n,k,s}^{\UE,t} (\bp[t]) \! \leq  \!
	R_n^{\BS,t} (\bP[t],\!\bW^\BS[t]), \forall (n,t), \subnum \\
\hspace{-5mm}	(\tilde{C}7):   \scaleobj{.8}{ \sum_{\forall (n,s)}} p_{n,k,s}[t] \leq p_k^{\max}, \forall (k,t). \subnum 
\end{IEEEeqnarray}
In addition, the arguments $\balpha[t]$ and $\bmu[t]$ can be omitted in corresponding functions, i.e., $R_{k}^{\UE,t}(\bp[t])$, $R_{n,k,s}^{\UE,t}(\bp[t]), R_n^{\BS,t}(\bP[t],\bW^\BS[t])$ and $R_{m,n}^{\BS,t}(\bP[t],\bW^\BS[t])$.
As a result, the equivalent problem at iteration $i$ of \eqref{eq: Max rate} at TS $t$ can be formulated as
\beq\label{eq: Max rate 2}
\max_{\bp,\bP,\bW^\BS}  \scaleobj{.8}{ \sum_{\forall (k,n,s)}} \omega_k[t] R_{n,k,s}^{\UE,t} (\bp[t])	\text{ s.t. }    (\tilde{C}1)-(\tilde{C}7),(C8).  
\eeq
The load balancing constraint $(\tilde{C}6)$ can be transformed into a more traceable form by the following theorem.
\begin{theorem}
	The solution for \eqref{eq: Max rate 2} can be obtained by solving the following problem, which has the same optimal solution with \eqref{eq: Max rate 2} as
	\bIEEEeq{cl}\label{eq: Max rate 3}
	\hspace{-5mm}\max_{\bp,\bP,\bW^\BS, \blambda^\UE, \blambda^\BS}  \quad & 	{\scaleobj{.8}{\sum_{\forall (k,n)}}} \omega_k[t] \lambda_{n,k}^\UE[t] \subnum \label{eq: Max rate, power and BW 2 a}\\
	\hspace{-20mm} \st     & \hspace{-15mm} \text{constraints } (\tilde{C}1) - (\tilde{C}7), (C8),  \nonumber \\
	& \hspace{-15mm} (C 9): \scaleobj{.8}{\sum_{\forall k \in \calK}}\lambda_{n,k}^\UE[t] \leq 
	\lambda_{n}^\BS[t], \forall (n,t), \subnum \label{eq: Max rate 3 b} \\
	& \hspace{-15mm} (C10): \lambda_{n,k}^\UE[t] \leq \scaleobj{.8}{\sum_{\forall s \in \calNsc}} R_{n,k,s}^{\UE,t} (\bp[t]), \forall (n,k) ,  \subnum \label{eq: Max rate 3 c} \\
	& \hspace{-15mm} (C11): \lambda_n^\BS[t] \leq \scaleobj{.8}{\sum_{\forall m \in \calM} } R_{m,n}^{\BS,t} (\bP[t],\bW^\BS[t]),  \forall n, \subnum \label{eq: Max rate 3 d}
	\eIEEEeq
	where $\blambda^\UE \triangleq \{\lambda_{n,k}^\UE [t] \}_{\forall (n,k,t)}$ and $\blambda^\BS \triangleq \{ \lambda_{n}^\BS[t] \}_{\forall (n,t)}$ are new variables, which are introduced as a lower bound of the UE and BS transmission data functions, respectively.
\end{theorem}
\begin{IEEEproof}
Due to the lack of space, the proof can be given simply as follows.	
 At each TS, each BS receives the data amount from associated UEs and forwards them to the serving LEO satellite. Therefore, at the optimal point of problem \eqref{eq: Max rate 2}, the total data throughput from UEs to $\BS_{n}$ and that from $\BS_{n}$ to LEO satellites must be equal, i.e., constraints $(\tilde{C}6)$ holds with equality. As a result, constraint $(\tilde{C}6)$ can be replaced by $(C9)-(C11)$. In addition, exploiting the condition given in $(C10)$, the objective function \eqref{eq: Max rate 3} can be replaced by that of \eqref{eq: Max rate, power and BW 2 a} without changing the optimal solution. 
\end{IEEEproof}
	
Problem \eqref{eq: Max rate 3} is still non-convex because of the non-convexity of constraint $(C10)$. To convexify $(C10)$, the transmission data function of UE is approximated and transformed into a convex form, which is described in the next subsection.

\subsection{Successive Convex Approx. for Low-complExity (SCALE)}
In this subsection, SCALE method is employed to convexify constraint $ (C10) $. First, to address the DC component $ 1 $ in logarithmic function $ \log(1+x) $, we use the following iterative lower bound
$a^{(i)} \log(x) + b^{(i)} \leq \log(1+x)$,
where $a^{(i)}$ and $b^{(i)}$ can be updated as $a^{(i)} = {x^{(i-1)}}/{(x^{(i-1)}+1)}, b^{(i)} = \log(1+x^{(i-1)}) - a^{(i)} \log(x^{(i-1)})$ \cite{SCALE,TamTran_VTC16}.
Accordingly, the UE transmission rate $ R^{\UE,t}_{n,k,s} (\bp[t]) $ has a lower bound as
\beq \label{eq: lower bound R_UE SCALE}
	R_{n,k,s}^{\UE,t} (\bp[t]) \geq T_S \Wsc \Big(a_{n,k,s}^{(i)}\log_2\big(\mathtt{SINR}_{n,k,s}^t\big) \! + \! b_{n,k,s}^{(i)}\Big)
\eeq
Subsequently, employing the new variable $\bar{p}_{n,k,s}[t]$'s which satisfies $ p_{n,k,s}[t] = \exp(\bar{p}_{n,k,s}[t]), \; \forall (n,k,s) $ \cite{TiNguyen_TWC20}, the RHS of \eqref{eq: lower bound R_UE SCALE} can be naturally transformed to a concave function \cite{SCALE}. Hence, $ (C10) $ can be rewritten in a convex form as
\begin{IEEEeqnarray}{ll} \label{eq:C10_tilde}
	& (\tilde{C}10): \lambda_{n,k}^\UE[t] \leq T_S \Wsc \scaleobj{.8}{\sum_{\forall s \in \calNsc}} \!\!\! \Big(a_{n,k,s}^{(i)}\big( \log_2 h_{n,k,s}[t] + \bar{p}_{n,k,s}[t] \non \\
	& \quad - \log_2 \big(\scaleobj{.8}{\sum_{k' \neq k}}{ h_{n,k,s}[t] \exp(\bar{p}_{n,k',s}[t]) } + \sigma_n^2 \big) \big) + b_{n,k,s}^{(i)} \Big).
\end{IEEEeqnarray}
In addition, utilizing $ p_{n,k,s}[t] = \exp(\bar{p}_{n,k,s}[t]) $ \cite{TiNguyen_TWC20}, one can rewrite $(\tilde{C}1)$-$(\tilde{C}3)$ and $(\tilde{C}7)$ as
\begin{IEEEeqnarray}{ll}\label{eq:C1237_tilde}
	(\bar{C}1): \scaleobj{.8}{\sum_{\forall k \in \calK} }\zeta_{n,k,s}^{(i)}[t] \exp(\bar{p}_{n,k,s}[t]) \leq 1, \forall (n,s,t) , \subnum \\
	(\bar{C}2):  \scaleobj{.8}{\sum_{\forall s \in \calNsc}} \zeta_{n,k,s}^{(i)}[t] \exp(\bar{p}_{n,k,s}[t]) \leq \bar{S}, \forall (n,k,t), \subnum \\
	(\bar{C}3):  \scaleobj{.8}{\sum_{\forall n } }\xi_{n,k}^{(i)}[t] \sum_{\forall s} \exp(\bar{p}_{n,k,s}[t]) \leq 1, \forall (k,t), \subnum \\
	(\bar{C}7):  \scaleobj{.8}{\sum_{\forall (n,s)} } \exp(\bar{p}_{n,k,s}[t]) \leq p_k^{\max}, \forall (k,t). \subnum 
\end{IEEEeqnarray}
It is worth noting that these constraints are convex because their left-hand-side functions are log-sum-exp forms.
Using the approximation results given in \eqref{eq: lower bound R_UE SCALE}, \eqref{eq:C10_tilde}, and \eqref{eq:C1237_tilde}, we can rewrite problem \eqref{eq: Max rate 3} as
\bIEEEeq{cl}\label{eq: Max rate 4}
    \max_{\bar{\bp},\bP,\bW^\BS, \blambda^\UE, \blambda^\BS}  \quad & 	{\scaleobj{.8}{\sum_{\forall (k,n)}}} \omega_k[t] \lambda_{n,k}^\UE[t] \label{eq: Max rate, power and BW 2 a}\\
  \hspace{-20mm}  \st     & \hspace{-20mm} (\bar{C}1) \!-\! (\bar{C}3),(\tilde{C}4),(\tilde{C}5),(\bar{C}7), (C8),(C9),(\tilde{C}10),(C11),  \nonumber 
\eIEEEeq
where $\bar{\bp} \triangleq [\bar{\bp}[t]]_{\forall t }$ and $\bar{\bp}[t] \triangleq [\bar{p}_{n,k,s}[t]]_{\forall (n,k,s) }$. 

Thanks to the CS and SCALE methods presented in the previous sections, problem \eqref{eq: Min time} can be addressed efficiently by solving problem \eqref{eq: Max rate 4} iteratively at each TS until the remaining data demand of all UEs is empty. 
The proposed algorithm is summarized as in Algorithm~\ref{alg_1}.
In addition, $\balpha$ and $\bmu$ can be rounded to binaries by using the method discussed in \cite{VuHa_TVT16} as
\beqn \label{eq: binary recovery}
   && \hspace{-10mm} \text{if } \zeta_{n,k,s}[t] p_{n,k,s}[t] \! \geq \! 1/2, \alpha_{n,k,s}[t] \!=\! 1, \text{ else }  \alpha_{n,k,s}[t] \! = \!0,  \subnum \\
   && \hspace{-10mm}  \text{if }  \xi_{m,n}[t] W^\BS_{m,n}[t] \! \geq \! 1/2, \mu_{m,n}[t] \!=\!
        1, \text{ else } \mu_{m,n}[t] \! = \!0. \subnum
\eeqn

\begin{algorithm}[t]
\footnotesize
	\begin{algorithmic}[1]
		\protect\caption{\textsc{Proposed Iterative Algorithm}}
		\label{alg_1}
        \STATE Set $t=1$.\\
        \WHILE {$t \leq N_T$ or $d_k[t]=0, \forall k$}
        \STATE Set $i=0, a_{n,k,s}^{(0)}=1, b_{n,k,s}^{(0)}=0, \; \forall (n,k,s)$, and generate an initial starting point $\Big(\bp^{(0)}[t], (W^\BS[t])^{(0)} \Big)$.\\
		\REPEAT
		\STATE Solve \eqref{eq: Max rate 4} to obtain $\Big(\bar{\bp}^\star[t], \bP^\star[t], (\bW^\BS[t])^\star\Big)$.
		\STATE Update\ \ $ \Big(\bar{\bp}^{(i+1)[t]}, (\bW^\BS[t])^{(i+1)}\Big) := \Big(\bar{\bp}^\star[t], (\bW^\BS[t])^\star \Big)$.
		\STATE Set $i=i+1$.
        \STATE Calculate $a_{n,k,s}^{(i)}, b_{n,k,s}^{(i)}$ and $\bp^{(i)}[t] = \exp(\bar{\bp}^{(i)}[t])$.
		\UNTIL Convergence
        \STATE Calculate $d_k[t], \forall k$ based on \eqref{eq: data remain}.
        \STATE Set $t=t+1$.
		\ENDWHILE
        \STATE Recovery association variables $\balpha$ and $\bmu$ by \eqref{eq: binary recovery}.
        \STATE \textbf{Output:} The solution $\Big(\bp^\star, \bP^\star, (\bW^\BS)^\star, \balpha^\star, \bmu^\star \Big)$.
    \end{algorithmic}
    \normalsize 
\end{algorithm}

\subsection{Greedy-Based Algorithm (GA)}
In order to perform a comparison of performance, we introduce a greedy algorithm (GA) in this section. For the transmission from BS to LEO satellite, each BS selects the LEO satellite with the best channel gain for forwarding the data at each time slot (TS). The LEO satellite then allocates the same bandwidth for all connected BSs and each BS transmits at maximum power. For the UE to BS transmission, let $\calK_{\mathtt{D}}$ be the set of UEs with remaining data demand. At each TS, UEs in $\calK_{\mathtt{D}}$ choose the BS with the best average channel gain to be served. Within each BS, SCs are assigned to UEs in descending order of channel gain. For simplicity, transmit power among SCs at UEs is allocated using a water-filling algorithm while ignoring inter-BS interference. However, due to the limited capacity of the BS-LEO satellite backhaul link, the auxiliary maximum power value at served UEs used for the water-filling algorithm is adjusted at each BS until the aggregate data rate of UEs is close to and lower than the BS rate. Let $R^\BS_n$ and $R^{\sum}_n$ denote the data rate of $\BS_n$ and the aggregate data rate without interference from UEs linked with $\BS_n$, respectively. In summary, GA is outlined in Algorithm~\ref{alg_2}.

\begin{algorithm}[t]
\footnotesize
\begin{algorithmic}[1]
    \protect\caption{\textsc{Greedy-Based Algorithm}}
    \label{alg_2}
    \STATE Set $t=1$ and $\calK_{\mathtt{D}} = \calK$.\\
    \WHILE {$t \leq N_T$ or $\calK_{\mathtt{D}} = \emptyset$}
    \FOR{$n=1 \rightarrow N$}
        \STATE $\BS_n$ associates with $\LEO_m$ satisfying $g_{m,n}[t]>g_{m',n}[t], \forall m' \neq m$. Set $\mu_{m,n}=1$.
        \STATE Set $P_{n}[t]:=P_n^{\max}$.
    \ENDFOR
    \STATE Each LEO satellite uniformly allocates bandwidth to linked BSs.
    \FORALL{$k \in \calK_{\mathtt{D}}$}
        \STATE $\UE_k$ chooses $\BS_n$ with the best channel gain.
    \ENDFOR
    \STATE Each BS allocates SCs to its UEs in descending order of channel gain.
    \STATE Build corresponding matrix $\balpha$.
    \FOR{$n=1 \rightarrow N$}
    \STATE Set $p^{\sf{up}}=2 p^{\max}_\UE$ and $p^{\sf{low}} = 0$.
    \REPEAT
    \STATE Set $\bar{p}^{\max}_\UE = (p^{\sf{up}}+p^{\sf{low}})/2$.
    \STATE Utilize water-filling algorithm for each UE linked with $\BS_n$ to find power allocation using $\bar{p}^{\max}_\UE$ as the maximum power.
    \IF{$R^\BS_n < R^{\sum}_n$}
    \STATE Set $p^{\sf{up}} = \bar{p}^{\max}_\UE$.
    \ELSE
    \STATE Set $p^{\sf{low}} = \bar{p}^{\max}_\UE$.
    \ENDIF
    \UNTIL $R^\BS_n > R^{\sum}_n$ and $R^\BS_n - R^{\sum}_n \leq \epsilon$
    \ENDFOR
    \STATE Calculate $d_k[t], \forall k$ based on \eqref{eq: data remain}.
    \IF{There exists $\UE_k$ in $\calK_{\mathtt{D}}$ with $d_k[t]=0$}
    \STATE $\calK_{\mathtt{D}} = \calK_{\mathtt{D}} - \{k\}$.
    \ENDIF
    \STATE Set $t=t+1$.
    \ENDWHILE
    \STATE \textbf{Output:} The solution $(\bp, \bP, \bW^\BS, \balpha,\bmu )$.
\end{algorithmic}
 \normalsize 
\end{algorithm}

\section{Numerical Results}
This section presents numerical results to assess the efficacy of the proposed algorithms and examine the influence of various parameters. The simulations were carried out in an area with dimensions of $5$km $\times$ $6$km located at geographical coordinates $(40^\circ \text{N}, 20^\circ \text{E})$ and comprised of $N$ BSs and $K$ UEs. The BSs were deployed in clusters, each consisting of 3 BSs and serving 4 UEs. Three LEO satellites were used to serve the terrestrial network, located at $(\varphi_{1},\theta_{1}) = (39.93^\circ \text{N}, 19.99^\circ \text{E}), (\varphi_{2},\theta_{2}) = (39.97^\circ \text{N}, 19.99^\circ \text{E})$ and $(\varphi_{3},\theta_{3}) = (39.95^\circ \text{N}, 20.03^\circ \text{E})$ at TS 1. The key parameters are listed in Table~\ref{tab:parameter}. The LEO beam pattern is defined as per \cite{3gpp.38.811}. The channel model of link BS-UE was assumed to be Rician channel with path-loss $\mathtt{PL_{\BS,\UE}}=145.4+37.5 \log(d_{\BS,\UE})$, while the channel model for link LEO satellite-BS was used as per \cite{Hung_WSA23,VuHaGC2022}.

\begin{table}[!t]
	\caption{Simulation Parameters}
  \vspace{-2mm}
	\label{tab:parameter}
	\centering
    \scalebox{0.8}{
	\begin{tabular}{l|l}
		\hline
		Parameter & Value \\
		\hline\hline
        LEO satellite bandwidth used for TN, $W^\LEO_{m} = W^\LEO,\; \forall m$   &  20 MHz \\
        LEO satellite altitude                  & 600 km \\
        BS-Satellite operation frequency, $f_c$              & 30 GHz \\
		Noise power density at BS and LEO satellite	& -174 dBm/Hz \\
		Maximum power at BS, $ P^{\max}_{n} = P^{\max}_\BS, \; \forall n$	& 14 dBW \\
		Number of UEs, $ K $					& 48 \\
        Number of BSs, $ N $					& 12 \\
        Number of BS clusters                   & 4 \\
        Number of visible LEO satellites, $M$           & 3 \\
		UE data demand, $ D_k, \; \forall k $	& 2.5 Mbits \\
        SC bandwidth, $\Wsc$                    & 360 kHz \\
        Number of SCs, $N_SC$                   & 8 \\
        TS duration, $T_S$                      & 30 ms \\
        Number of considered TSs, $N_T$         & 50 \\
		\hline		   				
	\end{tabular}}
\end{table}

\begin{figure}[!t]
	\centering
	\includegraphics[width=0.88\linewidth]{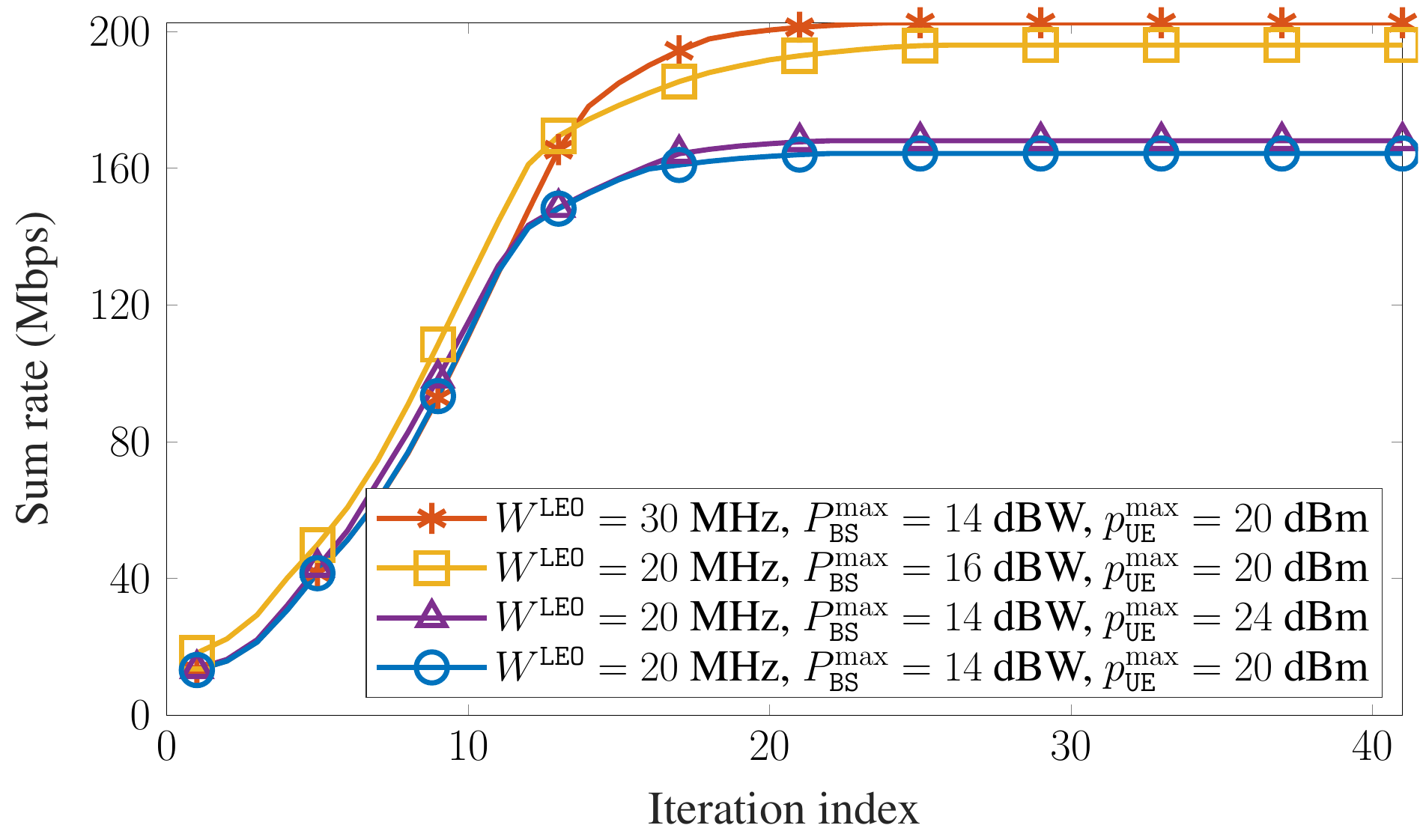}
  \vspace{-3mm}
	\caption{The sum rate convergence of Algorithm 1.}
	\label{fig:SR_Convergence}
  \vspace{-3mm}
\end{figure}
Fig.~\ref{fig:SR_Convergence} shows the convergence of the sum rate (SR) for Algorithm~\ref{alg_1} at TS $t=1$ under different combinations of the LEO satellite bandwidth and maximum power of BSs and UEs. It can be observed that in all considered scenarios, the SR exhibits a similar trend where it increases rapidly and reaches its saturation value after a few tens of iterations which has confirmed the convergence of Algorithm~\ref{alg_1}. For example, our proposed approach converges after approximately 20 iterations when the parameters are $(W^\LEO,P^{\max}_\BS,p^{\max}_\UE)=(20,14,24)$ or $(20,14,20)$, and after around 25 iterations when $(W^\LEO,P^{\max}_\BS,p^{\max}_\UE)=(30,14,20)$ or $(20,16,20)$.

\begin{figure}[!t]
	\centering
	\includegraphics[width=0.88\linewidth]{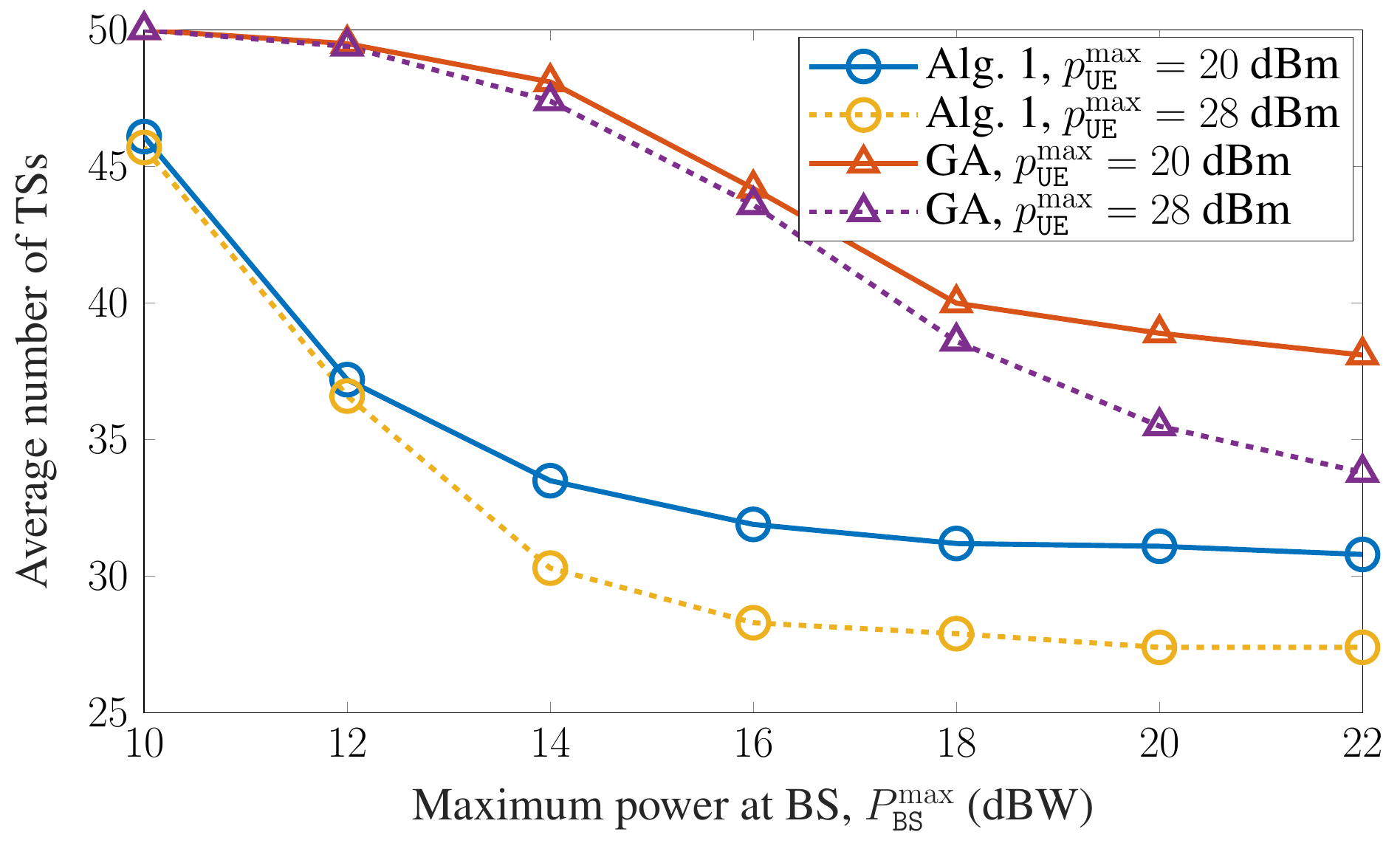}
  \vspace{-3mm}
	\caption{The average number of required TSs versus BSs' maximum power.}
	\label{fig:NumTS_PBS}
  \vspace{-3mm}
\end{figure}

Fig.~\ref{fig:NumTS_PBS} shows the number of TSs needed to meet the UE data demand versus the maximum power at BS. It can be seen that the required transmission time decreases with an increase in maximum power at BS. Our proposed framework can satisfy all users in a shorter time duration than GA does where a significant gap between these two solutions can be seen in this figure. 
Interestingly, the gap separating the required TS numbers of our proposed algorithm corresponding to $p^{\max}_{\UE}=20$ and $28$ dBm is very small when $P^{\max}_\BS \in [10,12]$ dBW, while that becomes larger when $P^{\max}_\BS$ gets higher, i.e., about 3.5 TSs. This implies that the low $P^{\max}_\BS$ limits backhaul link capacity and results in a bottleneck. However, that gap corresponding GA is quite small over all range of $P^{\max}_\BS$. This has shown the efficiency of Algorithm~\ref{alg_1} since it can allocate resources dynamically to achieve better backhaul link capacity compared with GA. Furthermore, Algorithm~\ref{alg_1} can satisfy the UE data demand within the considered time at all examined values of $P^{\max}_{\BS}$, however, at $P^{\max}_{\BS} = 10$ dBW GA can not ensure the UE data demand, i.e., the average remaining data demand at this point is about $0.88$ Mbits in both cases of $p^{\max}_{\UE}$.

\begin{figure}[!t]
	\centering
	\includegraphics[width=0.88\linewidth]{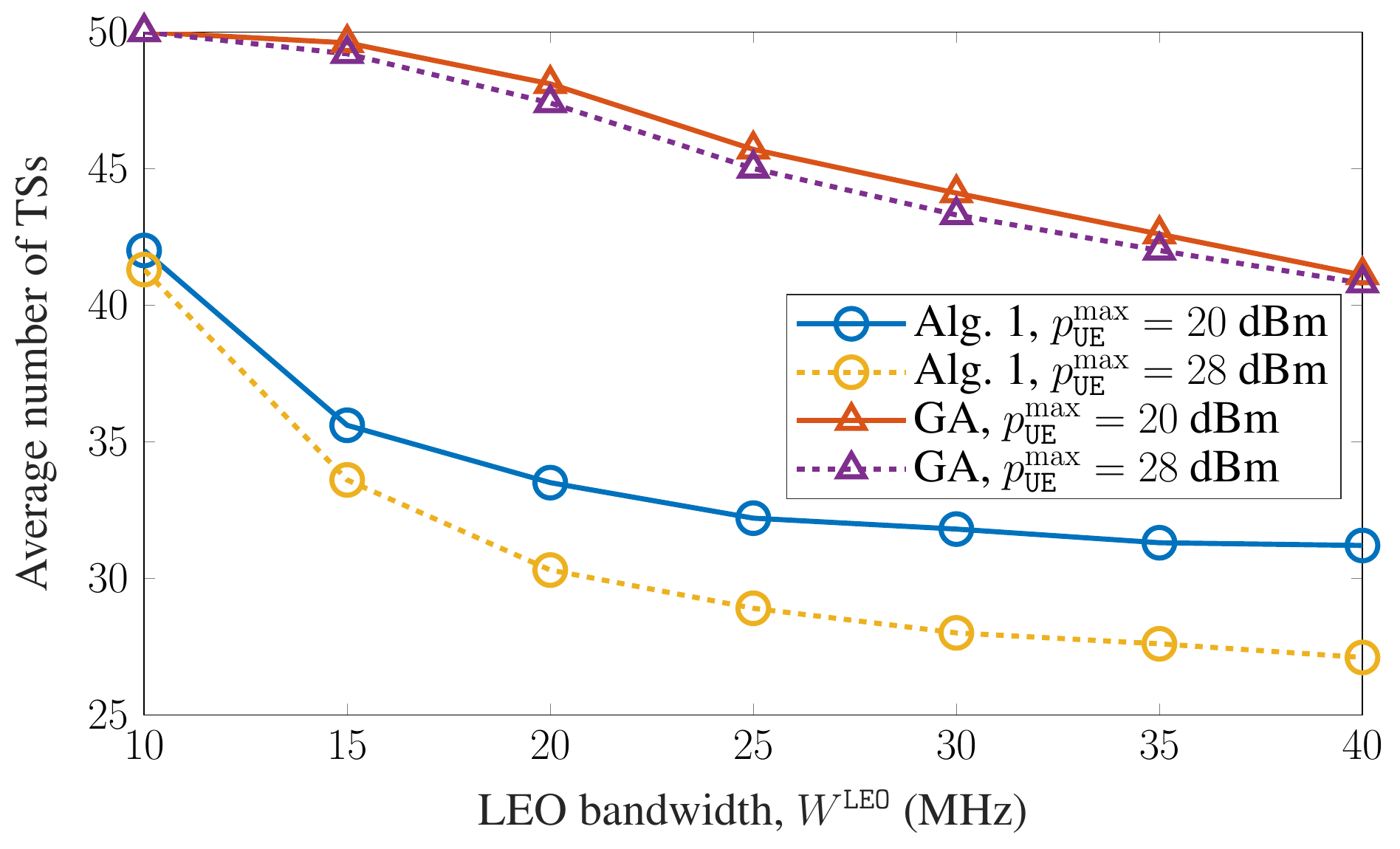}
  \vspace{-3mm}
	\caption{The average number of required TSs versus the LEO bandwidth.}
	\label{fig:NumTS_WLEO}
  \vspace{-3mm}
\end{figure}
Instead of changing the maximum power at BS, the impact of the change in LEO satellite bandwidth on the number of required TSs is shown as in Fig.~\ref{fig:NumTS_WLEO}. 
As expected, when $W^\LEO$ increases, both approaches can reduce the transmission time significantly.
It can be seen that Algorithm 1 outperforms the GA significantly in terms of achieving a lower average number of required TSs to complete the data demand for all UEs. In particular, those numbers of the proposed and greedy algorithms at $W^\LEO = 20$ MHz are about $33.5$ and $48.1$ TSs when $p^{\max}_{\UE}=20$ dBm, and $30.3$ and $47.4$ TSs when $p^{\max}_{\UE}=28$ dBm, respectively.
For GA, the increasing of $p^{\max}_{\UE}$ from $20$ dBm to $28$ dBm does not improve significantly the performance, this indicates that there exists a backhaul-link bottleneck. Whereas for Algorithm~\ref{alg_1}, one shows that the higher $p^{\max}_{\UE}$ is allocated, the shorter transmission time can be achieved.
Furthermore, similar to Fig.~\ref{fig:NumTS_PBS}, Algorithm 1 can satisfy the entire UE data demand at all considered points of $W^\LEO$, whereas at $W^\LEO=10$ MHz GA can not complete the data demand for all UEs.

\begin{figure}
	\centering
	\includegraphics[width=0.88\linewidth]{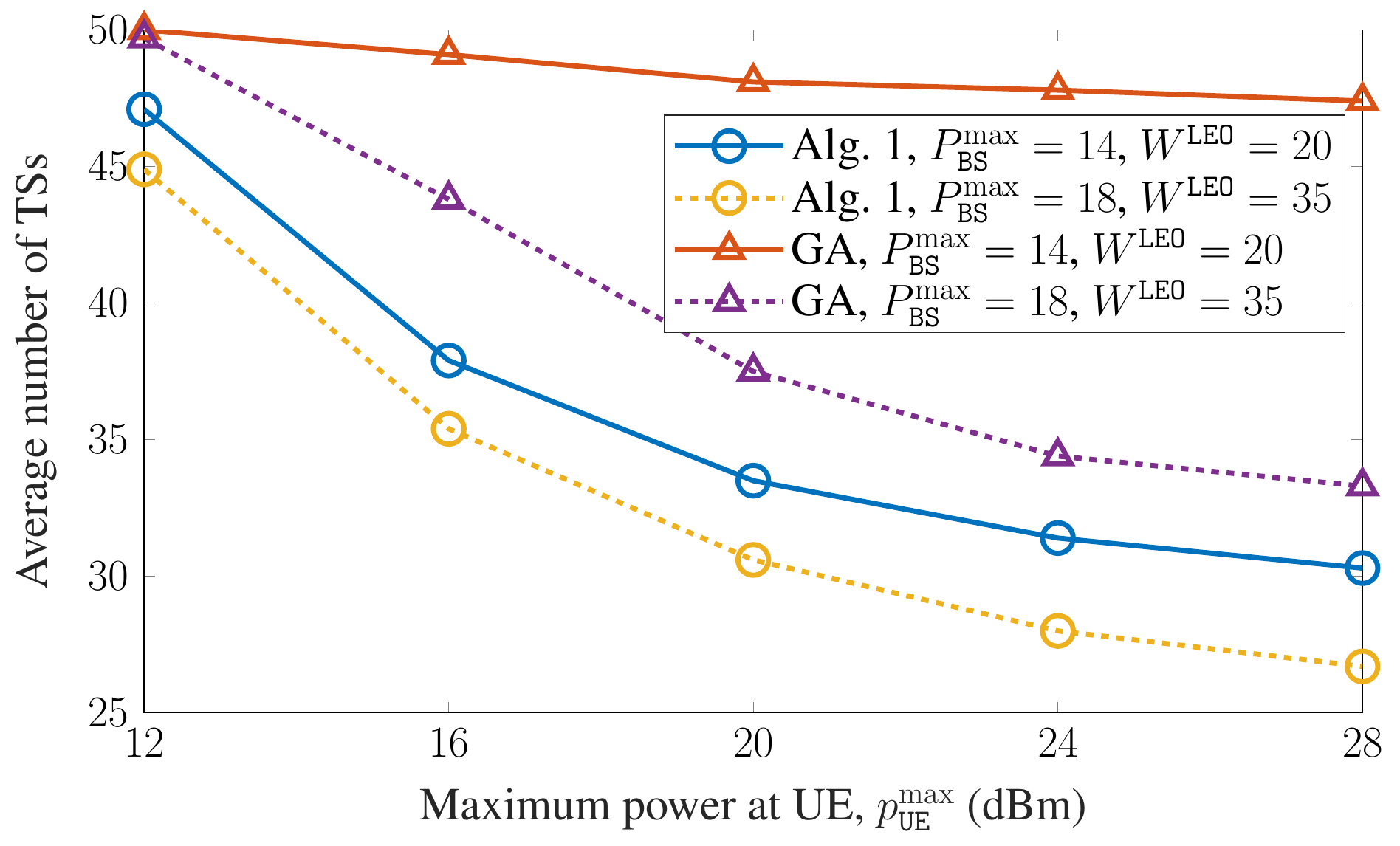}
  \vspace{-3mm}
	\caption{The average number of required TSs versus UEs' maximum power.}
	\label{fig:NumTS_pUE}
  \vspace{-3mm}
\end{figure}
Fig.~\ref{fig:NumTS_pUE} presents the average number of required TSs as a function of the maximum power at UE. In general, the average number of required TSs decreases when $p^{\max}_\UE$ increases. For GA, the performance is insignificantly improved with higher $p^{\max}_\UE$ in case $(P^{\max}_\BS,W^\LEO)=(14,20)$; however, the average number of required TSs decrease quickly as $p^{\max}_\UE$ increases in case $(P^{\max}_\BS,W^\LEO)=(18,35)$. This improvement is due to the increasing of backhaul link capacity owing to the raising of both $P^{\max}_\BS$ and $W^\LEO$. As expected, Algorithm 1 outperforms GA in terms of the lower average number of transmission TSs, even when comparing Algorithm in case $(P^{\max}_\BS,W^\LEO)=(18,35)$ and GA in case $(P^{\max}_\BS,W^\LEO)=(14,20)$. Specifically, GA and Algorithm 1 require $47.8$ and $31.4$ with $(P^{\max}_\BS,W^\LEO)=(14,20)$, and $34.4$ and $28$ TSs with $(P^{\max}_\BS,W^\LEO)=(18,35)$ to satisfy UE data demand at $p^{\max}_\UE=24$ dBm. Thus this figure further shows the outperformance of Algorithm 1 in terms of minimizing the number of transmission TSs but also satisfying the UE data demand in all examined scenarios.

\section{Conclusion}
In conclusion, this paper has presented a study of the design of a two-tier user association and resource allocation for an integrated satellite-terrestrial network that utilizes Low-Earth-Orbit (LEO) satellites as a backhaul link. The optimization problem of satellite-BS-UE association and resource allocation under the constraints of load balance and UE demand was addressed through the development of an iterative algorithm based on approximation and relaxation methods. The results of this study provide valuable insights into the design and optimization of integrated satellite-terrestrial networks and can help to advance the development of next-generation communication systems.





\section*{Acknowledgment}
This work has been supported by the Luxembourg National Research Fund (FNR) under the project INSTRUCT (IPBG19/14016225/INSTRUCT) and project MegaLEO (C20/IS/14767486).

\bibliographystyle{IEEEtran}
\bibliography{Journal}

\begin{thebibliography}{10}
\providecommand{\url}[1]{#1}
\csname url@samestyle\endcsname
\providecommand{\newblock}{\relax}
\providecommand{\bibinfo}[2]{#2}
\providecommand{\BIBentrySTDinterwordspacing}{\spaceskip=0pt\relax}
\providecommand{\BIBentryALTinterwordstretchfactor}{4}
\providecommand{\BIBentryALTinterwordspacing}{\spaceskip=\fontdimen2\font plus
\BIBentryALTinterwordstretchfactor\fontdimen3\font minus
  \fontdimen4\font\relax}
\providecommand{\BIBforeignlanguage}[2]{{%
\expandafter\ifx\csname l@#1\endcsname\relax
\typeout{** WARNING: IEEEtran.bst: No hyphenation pattern has been}%
\typeout{** loaded for the language `#1'. Using the pattern for}%
\typeout{** the default language instead.}%
\else
\language=\csname l@#1\endcsname
\fi
#2}}
\providecommand{\BIBdecl}{\relax}
\BIBdecl

\bibitem{SatCom_survey_and_challenge}
O.~Kodheli, E.~Lagunas, N.~Maturo, S.~K. Sharma, B.~Shankar, J.~F.~M. Montoya,
  J.~C.~M. Duncan, D.~Spano, S.~Chatzinotas, S.~Kisseleff, J.~Querol, L.~Lei,
  T.~X. Vu, and G.~Goussetis, ``Satellite communications in the new space era:
  A survey and future challenges,'' \emph{IEEE Commun. Surveys Tut.}, vol.~23,
  no.~1, pp. 70--109, 1st Quart. 2021.

\bibitem{Servey_NGSO}
H.~Al-Hraishawi, H.~Chougrani, S.~Kisseleff, E.~Lagunas, and S.~Chatzinotas,
  ``A survey on non-geostationary satellite systems: The communication
  perspective,'' \emph{IEEE Commun. Surveys Tut.}, Aug. 2022.

\bibitem{VuHa_ICC23}
V.~N. Ha, E.~Lagunas, T.~S. Abdu, H.~Chaker, S.~Chatzinotas, and J.~Grotz,
  ``Large-scale beam placement and resource allocation design for
  {MEO}-constellation {SATCOM},'' in \emph{ICC Workshop - 6GSatComNet}, 2023.

\bibitem{ISTN_Toward_6G_App_challenge}
X.~Zhu and C.~Jiang, ``Integrated satellite-terrestrial networks toward {6G}:
  Architectures, applications, and challenges,'' \emph{IEEE Internet Things
  J.}, vol.~9, no.~1, pp. 437--461, Jan. 2022.

\bibitem{Tedros_ICC23}
T.~S. Abdu, E.~Lagunas, V.~N. Ha, J.~Grotz, S.~Kisseleff, and S.~Chatzinotas,
  ``Demand-aware flexible handover strategy for leo constellation,'' in
  \emph{ICC Workshop - 6GSatComNet}, 2023.

\bibitem{twotier_UE_HAP_LEO_association}
L.~Zhang, H.~Zhang, C.~Guo, H.~Xu, L.~Song, and Z.~Han, ``Satellite-aerial
  integrated computing in disasters: User association and offloading
  decision,'' in \emph{ICC 2020 - 2020 IEEE Inter. Conf. Commun. (ICC)}, June
  2020, pp. 554--559.

\bibitem{LEO_HO_Bipartite_GamePotential}
Y.~Wu, G.~Hu, F.~Jin, and J.~Zu, ``A satellite handover strategy based on the
  potential game in {LEO} satellite networks,'' \emph{IEEE Access}, vol.~7, pp.
  133\,641--133\,652, Sept. 2019.

\bibitem{LEO_5G_Offloading}
B.~Di, H.~Zhang, L.~Song, Y.~Li, and G.~Y. Li, ``Ultra-dense {LEO}: Integrating
  terrestrial-satellite networks into {5G} and beyond for data offloading,''
  \emph{IEEE Trans. Wireless Commun.}, vol.~18, no.~1, pp. 47--62, 2019.

\bibitem{LEO_IoT_MinTime}
Z.~Gao, A.~Liu, C.~Han, and X.~Liang, ``Max completion time optimization for
  internet of things in {LEO} satellite-terrestrial integrated networks,''
  \emph{IEEE Internet Things J.}, vol.~8, no.~12, pp. 9981--9994, 2021.

\bibitem{Hung_WSA23}
H.~Nguyen-Kha, V.~N. Ha, E.~Lagunas, S.~Chatzinotas, and J.~Grotz,
  ``{LEO}-to-user assignment and resource allocation for uplink transmit power
  minimization,'' in \emph{Proc. WSA \& SCC 2023}, 2023.

\bibitem{UltraDenseLEO_Offload_TN}
R.~Deng, B.~Di, S.~Chen, S.~Sun, and L.~Song, ``Ultra-dense {LEO} satellite
  offloading for terrestrial networks: How much to pay the satellite
  operator?'' \emph{IEEE Trans. Wireless Commun.}, vol.~19, no.~10, pp.
  6240--6254, Oct. 2020.

\bibitem{Min_Time_To_Weight_Max_SR}
F.~She, H.~Luo, W.~Chen, and X.~Wang, ``Joint queue control and user scheduling
  in {MIMO} broadcast channel under zero-forcing multiplexing,'' in \emph{2008
  IEEE International Conference on Communications}, May 2008, pp. 275--279.

\bibitem{compress_sensing}
E.~Candes, M.~Wakin, and S.~Boyd, ``Enhancing sparsity by reweighted {L1}
  minimization,'' \emph{J. Fourier Analysis and Applications}, vol.~14, pp.
  877--905, Dec. 2008.

\bibitem{VuHa_TGCN20}
V.~N. Ha, D.~H.~N. Nguyen, and J.-F. Frigon, ``System energy-efficient hybrid
  beamforming for mmwave multi-user systems,'' \emph{IEEE Trans. Green Commun.
  and Net.}, vol.~4, no.~4, pp. 1010--1023, 2020.

\bibitem{SCALE}
J.~Papandriopoulos and J.~S. Evans, ``Low-complexity distributed algorithms for
  spectrum balancing in multi-user {DSL} networks,'' in \emph{2006 IEEE
  International Conf. Commun.}, vol.~7, June 2006, pp. 3270--3275.

\bibitem{TamTran_VTC16}
T.~T. Tran, V.~N. Ha, L.~B. Le, and A.~Girard, ``Dynamic resource allocation
  for full-duplex {OFDMA} wireless cellular networks,'' in \emph{2016 IEEE 84th
  Vehicular Technology Conference (VTC-Fall)}, 2016, pp. 1--5.

\bibitem{TiNguyen_TWC20}
T.~T. Nguyen, V.~N. Ha, L.~B. Le, and R.~Schober, ``Joint data compression and
  computation offloading in hierarchical fog-cloud systems,'' \emph{IEEE Trans.
  Wireless Commun.}, vol.~19, no.~1, pp. 293--309, 2020.

\bibitem{VuHa_TVT16}
V.~N. Ha, L.~B. Le, and N.-D. Dao, ``Coordinated multipoint transmission design
  for cloud-{RANs} with limited fronthaul capacity constraints,'' \emph{IEEE
  Trans. Veh. Technol.}, vol.~65, no.~9, pp. 7432--7447, 2016.

\bibitem{3gpp.38.811}
3GPP, ``{Study on New Radio (NR) to support non-terrestrial networks},'' {3rd
  Generation Partnership Project (3GPP)}, Technical report (TR) 38.811, Sept.
  2020, version 15.4.0.

\bibitem{VuHaGC2022}
V.~N. Ha, T.~T. Nguyen, E.~Lagunas, J.~C. Merlano~Duncan, and S.~Chatzinotas,
  ``{GEO} payload power minimization: Joint precoding and beam hopping
  design,'' in \emph{GLOBECOM 2022 - 2022 IEEE Global Commun. Conf.}, 2022, pp.
  6445--6450.

\end{thebibliography}
\end{document}